\documentclass[%
 reprint,
 amsmath,amssymb,
 aps,
]{revtex4-2}

\usepackage{graphicx}
\usepackage{dcolumn}
\usepackage{bm}

\usepackage{amsmath}
\usepackage{mathtools}
\usepackage{mhchem}
\usepackage{verbatim}
\usepackage{xcolor}
\usepackage{ulem}

\begin{document}

\preprint{APS/123-QED}

\title{Search axion-like particle with space-based optical lattice clock network}
\title{Searching for axion-like particles with space-based optical lattice clock networks}

\author{Yu-Song Cao}
 \email{caoyusong15@mails.ucas.ac.cn}
\affiliation{School of Physics and Optoelectronic Engineering, Beijing University of Technology, Beijing, China}
 
\author{Nick Houston}
\email{nhouston@bjut.edu.cn}
\affiliation{School of Physics and Optoelectronic Engineering, Beijing University of Technology, Beijing, China}

\author{YanXia Liu}%
 \email{yxliu-china@ynu.edu.cn}
\affiliation{%
 School of Physics and Astronomy, Yunnan University, Kunming 650091, PR China
}%

\author{QingZheng Lv}%
\email{qzlv@gscaep.ac.cn}
\affiliation{%
Graduate School, China Academy of Engineering Physics, Beijing 100193, China
}%


\author{She-Sheng Xue}
\email{xue@icra.it}
\affiliation{%
 ICRANet Piazzale della Repubblica, 10 -65122, Pescara, Italy
}%
\affiliation{Physics Department, Sapienza University of Rome, Rome, Italy}
\affiliation{INFN, Sezione di Perugia, Perugia, Italy}
\affiliation{ICTP-AP, University of Chinese Academy of Sciences, Beijing, China}

\author{Ding-Fang Zeng}
 \email{dfzeng@bjut.edu.cn}
\affiliation{
 School of Physics and Optoelectronic Engineering, Beijing University of Technology, Beijing, China
}%


\date{\today}

\begin{abstract}
Axion like particles (ALPs) are particularly well motivated candidates for dark matter, and physics beyond the Standard Model more generally. 
However, direct detection remains extremely challenging due to the weakness of their interactions. 
In this work, we propose to employ space-based optical lattice clock networks (SOLCN), originally targeted to probe the stochastic gravitational wave background (SGWB), to detect axion dark matter through cosmic birefringence via minor modifications to their laser links. 
Searching for cross-correlated signals then provides significant sensitivity to low mass ALPs as a result of their large correlation lengths. 
In this work we focus on ALPs in the mass range
$10^{-18}-10^{-14}\text{eV}$, finding that we can probe down to $g_{a\gamma}\geq 10^{-12}\,\text{GeV}^{-1}$. 
Moreover, we find that the SOLCN sensitivity to the ALP-photon coupling depends crucially on the frequency of the network link. 
With an $8$ GHz frequency link, we then have the potential to reach $g_{a\gamma}\geq10^{-16}\,\text{GeV}^{-1}$ in this region.
\end{abstract}

\maketitle


\section{Introduction}
    Decades of observation have accumulated a wealth of evidence pointing to the existence of dark matter (DM) in our universe, whose nature lies beyond the Standard Model. 
    According to the standard $\Lambda\text{CDM}$ cosmology, DM particles should be long-lived and interact weakly with standard model particles \cite{APJ261,APJ501,AA641,MNRAS249,MNRAS468}. 

    Among the candidates for DM particles, ALPs are of particular interest, as they can emerge in different scenarios via totally independent motivations. 
    More specifically, the axion was originally introduced by Peccei and Quinn as a solution to the strong CP problem in quantum chromodynamics as a minimal extension of the Standard Model \cite{L38,D16}. 
    Meanwhile, in string theory and supergravity, the necessity of compactification of extra dimensions also generically leads to the existence of particles with properties similar to those of the axion, generally referred to as ALPs. 
    Due to this diversity of theoretical underpinning, and several possible production mechanisms, the predicted ALP mass covers a broad range \cite{JHEP06,D81,Xue2022}. 
    However, cosmological observations suggest that for ALPs comprising DM the favored mass range is $m_{a}\ll 1\text{eV}$.

    Besides the theoretical motivations, ALPs are also widely studied experimentally since they couple to electromagnetism. 
    So far, many types of experiments have been conducted or proposed to probe the parameter space of the axion-photon coupling $g_{a\gamma}$ and mass $m_a$ of ALPs \cite{nick,PPNP102,D113-055038,D110-103525,D103,L126,epjc81,L121-241102,D113-083010,D113-043001,D113-023053,D112,L136,L135,D113-092004,D113-072011,D109,L121-161301,plb659,CAST,M87,SN,bhp,athena,dance,sgra,L124,Bi:2020ths}. 
    Due to the weakness of the coupling between ALPs and photons, it is natural to look to precision measurement techniques from other fields of physics, such as gravitational wave detection. \cite{D98,D111,D113-043059,2410,uop,L123}. 
    
    Detecting ALPs with gravitational wave detectors can be achieved via the cosmic birefringence effect, where left and right polarized photons traveling in the ALP DM halo will obtain different dispersion relations \cite{D41,L81,epjc84}. 
    Specifically, this birefringence effect can manifests as a rotation of the polarization plane in polarimetry \cite{D111,D113-043059}, a shift of the interference pattern in laser interferometry \cite{D98,2410}, or a frequency shift in laser links between optical lattice clocks \cite{uop}. 

    This is particularly noteworthy as the last decade has witnessed a surge of development toward detecting low-frequency gravitational waves with optical lattice clocks \cite{D94-12403,epjd73,epjd74,scpma,2602,2605,APJ991,QST7,CQG37}. 
    To perform detection, one needs two synchronised optical lattice clocks connected by an ultrastable laser link \cite{D94-12403}. Comparing the rates of the two clocks allows one to measure the frequency shift of the laser link with extreme accuracy. 
    In this work we propose to leverage these space-based optical lattice clock networks (SOLCN), originally proposed as an alternative scheme for detecting the stochastic gravitational wave background, for ALP DM detection \cite{scpma}. 
    
\begin{figure}
\begin{center}
\includegraphics[totalheight=45mm]{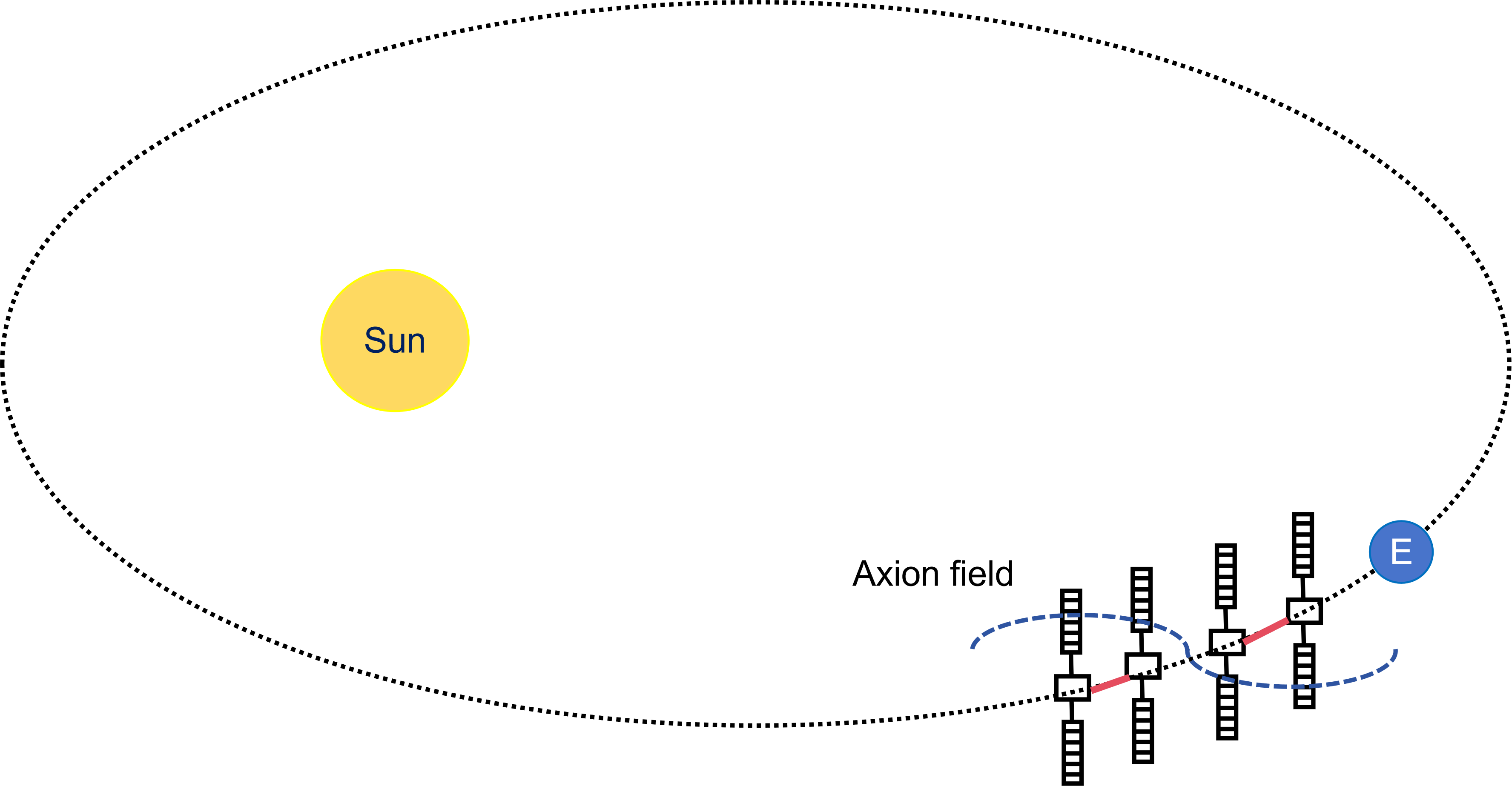}
\caption{Scheme of a SOLCN. We have four satellites arranged along Earth's orbit around the sun are grouped into two pairs. Within each pair, two satellites share a laser link with left polarization. The distance between the satellites belonging to the pair is $L=5\times 10^{10}$m. The distance between the two pairs are chosen as $d=5\times 10^{6}$m to ensure the independence of the noise \cite{scpma}. The length scale of the whole system is smaller than the coherence length of the axion field, resulting in the correlated signals in the two pairs.}
\label{fig:schem}
\end{center}
\end{figure}

    In simple terms, in SOLCN we have more than one pair of optical lattice clock connected by a laser link, as shown in Fig.(\ref{fig:schem}). 
    As we will show in detail later, the ALP DM halo will induce correlated signals among different pairs of optical lattice clocks since the ALPs field is coherent within the length scale of the SOLCN, while the noise between different optical lattice clocks is uncorrelated due to the large distance between them. 
    Searching for cross-correlated signals can thus provide good sensitivity to the underlying dark matter effects. 

    This paper is organized as follows. 
    In the next section, we will derive the dispersion relation of left and right-polarised photons in the presence of ALP DM. 
    Following that, we derive the corresponding signal and noise for a SOLCN, and then show the projected sensitivity. 
    Conclusion and discussions are given in closing. 
    Throughout this paper we use natural units where $\hbar=c=1$, with factors of $h$ and $c$ explicitly shown in certain formulae for clarity.

\section{Photon dispersion in the presence of ALP DM}
    In this section we firstly derive the cosmological birefringence induced by the ALP DM halo. 
    To begin we consider the Lagrangian for axion electrodynamics,
    \begin{equation}\label{eq:lagrangian}
        \mathcal{L}=-\frac{1}{4}F^{\mu\nu}F_{\mu\nu}-\frac{g_{a\gamma}}{4}\phi F^{\mu\nu}\tilde{F}_{\mu\nu},
    \end{equation}
    where $F^{\mu\nu}$ is the electromagnetic field strength tensor, $\phi$ is the axion field and $g_{a\gamma}$ is the axion-photon coupling strength. 
    Working in the gauge $A^{0}=0$, where $A^{0}$ is the scalar potential, Eq.\eqref{eq:lagrangian} gives the field equation for the vector potential as
    \begin{equation}\label{eq:fe}
        \nabla\times(\nabla\times\vec{A})+\frac{\partial^{2}\vec{A}}{\partial t^{2}}=-g_{a\gamma}(\nabla\times\vec{A})\frac{\partial\phi}{\partial t}.
    \end{equation}
    The background ALP field is given by
    \begin{equation}\label{eq:phi}
        \phi=\phi_{0}\cos{(m_{a}t+\theta)},
    \end{equation}
    where $m_{a}$ is the mass, $\theta$ is a phase factor and the amplitude $\phi_{0}$ is given by \cite{2410}
    \begin{equation}
        \phi_{0}=\frac{\sqrt{2\rho_{a}}}{m_{a}},
    \end{equation}
    where $\rho_{a}=0.45\text{GeV}/\text{cm}^{3}$ is the local DM density. 
    Note that in Eq.\eqref{eq:phi} we used the fact that the de-Broglie wavelength of ALPs within the mass range we are interested in is much larger than the characteristic length scale of the SOLCN we consider, and thus we can treat the ALP field as essentially homogeneous.

    Since the ALP mass under consideration is very small, the right-hand side of Eq.\eqref{eq:fe} should be a slowly varying function of time. 
    Thus, we can use the WKB approximation and write the solution of Eq.\eqref{eq:fe} as
    \begin{equation}\label{eq:mode}
        \vec{A}(t,\vec{x})=\int d\tilde{k}(A_{k}\nu_{k}(t)\vec{F}_{k}(t,\vec{x})+c.c.),
    \end{equation}
    where $d\tilde{k}$ is a covariant integral measure, and the time oscillating component can be written as
    \begin{equation}
        \nu_{k}(t)=e^{-i\int^{t}dt'\omega_{k}(t')},
    \end{equation}
    which by Eq.\eqref{eq:fe} leads to $\Ddot{\nu}_{k}/\nu_{k}=-\omega^{2}_{k}$.
    
    As the photon frequency here far exceeds the mass of the ALPs under consideration, the slowly varying envelope approximation demands \cite{PS98}
    \begin{equation}\label{eq:svea}
        \left|\frac{\Ddot{\vec{F}}_{k}}{\dot{\vec{F}}_{k}}\right|\ll\left|\frac{\dot{\vec{F}}_{k}}{\vec{F}_{k}}\right|\ll\omega_{k}.
    \end{equation}
    Substituting Eq.~\eqref{eq:mode}) into Eq.~\eqref{eq:fe}, we obtain the eigenvalue equation
    \begin{equation}\label{eq:eigenvalue}
        \nabla^{2}\vec{F}_{k}-g_{a\gamma}\frac{\partial\phi}{\partial t}(\nabla\times\vec{F}_{k})=-\omega_{k}^{2}\vec{F}_{k},
    \end{equation}
    where the `wavefunction' of the corresponding eigenstates is given by
    \begin{equation}\label{eq:circular}
        \vec{F}_{k,\pm}=\frac{1}{\sqrt{2}}(\hat{e}_{x}\pm i\hat{e}_{y})e^{-ikz},
    \end{equation}
    with eigenvalues
    \begin{equation}\label{eq:omegak}
        \omega_{k,\pm}=\sqrt{k^{2}\pm g_{a\gamma}\frac{\partial\phi}{\partial t}k}\simeq\omega_{0}\pm\frac{g_{a\gamma}}{2}\frac{\partial\phi}{\partial t},
    \end{equation}
    where $\omega_{0}=|k|$ is the dispersion relation of a photon in vacuum, and the term proportional to $g_{a\gamma}$ encodes the correction from the ALP DM halo.
    
\section{Searching for ALPs with SOLCN}
    In this section, we describe how we detect the ALP DM effect on the laser link frequency.
    The setup of the SOLCN is illustrated in Fig.(\ref{fig:schem}), where four satellites each carrying an optical lattice clock are arranged along Earth's orbit \cite{scpma}. 
    The four satellites are grouped into two pairs, and within each pair an ultrastable laser link is shared by the two satellites. 
    The length of the laser link between each pair is chosen as $L=5\times10^{10}\text{m}$ as in Ref.~\cite{scpma}. 
    Putting a quarter-wave plate in the light path of the laser link we obtain a circularly polarized laser beam, which is then sensitive to the cosmological birefringence discussed in the previous section. 
    We assume here this modification will not introduce an additional source of noise \cite{2410}, and for concreteness, we will assume that the laser links have left polarization in the following discussion. 
    During propagation, each photon in the laser link will experience a frequency shift as indicated in Eq.\eqref{eq:omegak}. 
    The two synchronized optical lattice clocks on the satellites will detect this frequency shift and so probe the ALP-derived signal.   
    
    The signal we measure is the ratio of the frequency shift of the laser link to its unshifted frequency
    \begin{equation}\label{eq:d}
        d(t)=\frac{\delta\omega_{k,+}}{\omega_{0}}=\frac{\omega_{k,+}(t_{d})}{\omega_{0}}-\frac{\omega_{k,+}(t_{e})}{\omega_{0}},
    \end{equation}
    where $t_{d}$ and $t_{e}$ are the times when the photon is detected and emitted. 
    Substituting Eq.\eqref{eq:omegak} into Eq.\eqref{eq:d}, we have
    \begin{equation}
        \begin{split}
            d(t)&=\frac{g_{a\gamma}\phi_{0}m_{a}}{2\omega_{0}}\left(\sin\left(m_{a}(t-L)+\varphi\right)-\sin\left(m_{a}t+\varphi\right)\right)\\
            &=-\frac{g_{a\gamma}\phi_{0}m_{a}}{\omega_{0}}\sin\left(\frac{m_{a}L}{2}\right)\cos\left(m_{a}t-\frac{m_{a}L}{2}+\varphi\right),
        \end{split}
    \end{equation}
    where we used the time at the detector $t_{d}=t$ as standard. 
    In the frequency domain, the signal $\tilde{d}(f)=\int dt e^{-i2\pi ft}d(t)$ reads
    \begin{equation}\label{eq:df}
        \begin{split}
            \tilde{d}(f)=&-\frac{g_{a\gamma}\phi_{0}m_{a}}{2\pi\omega_{0}}\sin\left(\frac{m_{a}L}{2}\right)\bigg(e^{-i(\frac{m_{a}L}{2}-\varphi)}\delta\left(f-\frac{m_{a}}{2\pi}\right)\\
            &+e^{i(\frac{m_{a}L}{2}-\varphi)}\delta\left(f+\frac{m_{a}}{2\pi}\right)\bigg).
        \end{split}
    \end{equation}

\begin{figure}
\begin{center}
\includegraphics[totalheight=50mm]{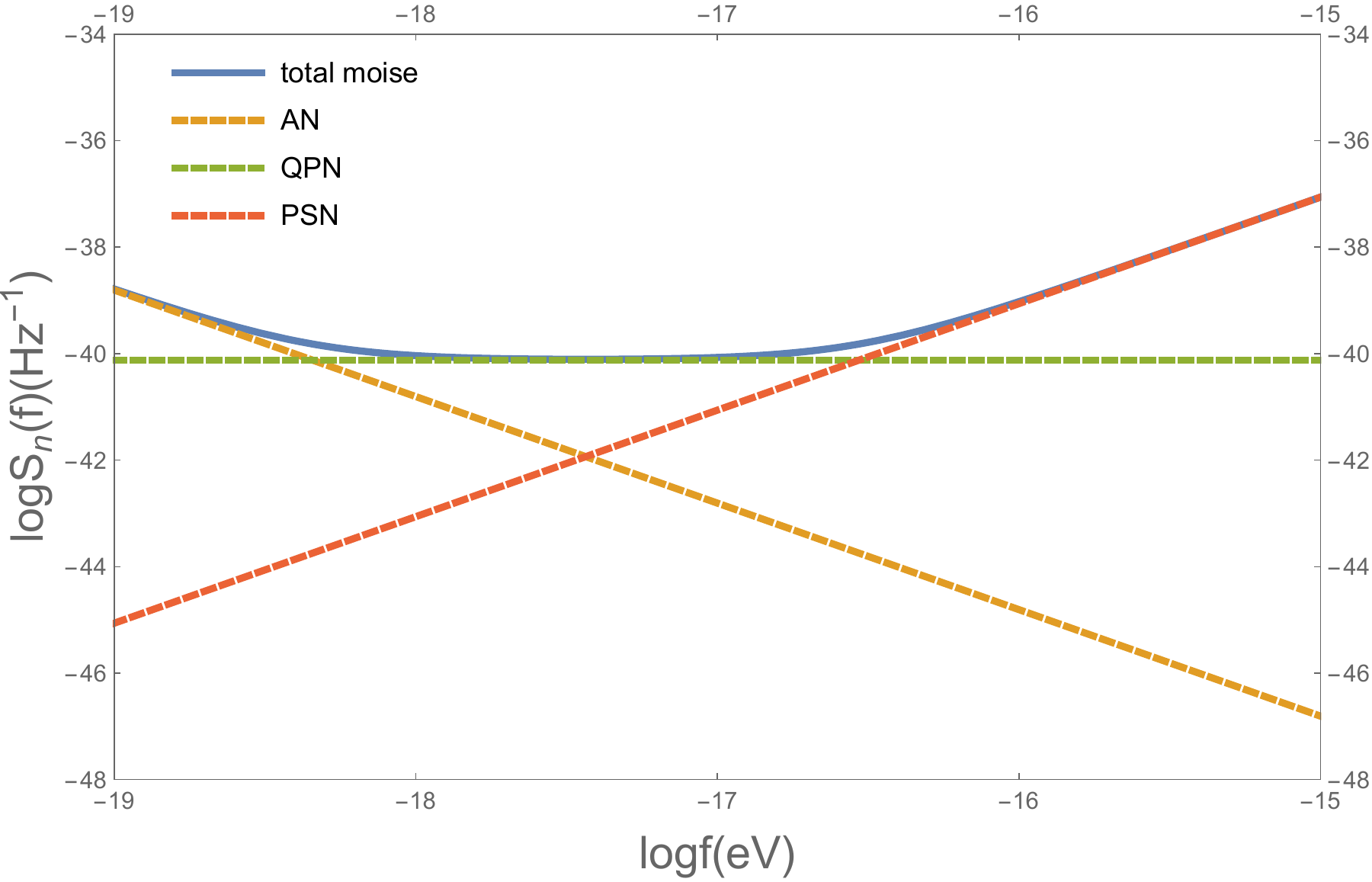}
\caption{Noise PSD of the SOLCN as a function of frequency. 
    The solid line is the total noise of the SOLCN, while the dashed lines are those for individual noise sources. 
    The total noise of the SOLCN reaches a minimum around $f\sim 10^{-17}-10^{-18}\text{eV}$.}
\label{fig:noise}
\end{center}
\end{figure}

    In a SOLCN, the noise comes mainly from three sources: acceleration noise (AN), quantum projection noise (QPN) and photon shot noise (PSN) \cite{scpma,D94-12403,epjd74}. 
    The AN noise comes from the station keeping and drag-free control of the satellites, and its power spectral density (PSD) is given by
    \begin{equation}
        S_{AN}(f)=\frac{S_{a}}{(2\pi f)^{2}c^{2}},
    \end{equation}
    where $S_{a}=9\times 10^{-30}\text{m}^{2}\cdot \text{s}^{-4}\cdot\text{Hz}^{-1}$ characterises the strength of AN noise \cite{epjd74}. 
    The QPN noise is a result of quantum measurement, specifically from fluctuations of the quantum states of the atoms in the optical lattice clocks. 
    The PSD for QPN noise can be written as \cite{scpma,D94-12403}
    \begin{equation}
        S_{QPN}(f)=\frac{2}{(2\pi f_{0})^{2}NT},
    \end{equation}
    where $T=160\text{s}$ is the Ramsey interrogation time, $N=7\times 10^{6}$ is the total number of atoms in the optical lattice clock and $f_{0}=\frac{\omega_{0}}{2\pi}=430\text{THz}$ is the transition frequency between the selected atomic states. 
    PSN noise is a consequence of the fluctuation in the measured arrival time of the laser photons, whose PSD is given by \cite{scpma,D94-12403}
    \begin{equation}
        S_{PSN}(f)=\frac{f^{2}(\Delta_{L}+\frac{h\nu_{0} B^{2}}{\eta P_{B}})}{\nu_{0}^{2}[(2\pi f)^{2}+B^{2}]},
    \end{equation}
    where $\eta=0.5$ is the quantum efficiency of the detector, $P_{B}=3\,\text{pW}$ is the power of the laser link transferred from one satellite to another, $\Delta_{L}=30\,\text{mHz}$ is the linewidth of the laser link, $B=\sqrt{\frac{\eta P_{B}\Delta_{L}}{h\nu_{0}}}$ and $h$ is Planck's constant. 
    The total noise PSD is then given by
    \begin{equation}\label{eq:sn}
        S_{n}(f)=S_{AN}(f)+S_{QPN}(f)+S_{PSN}(f).
    \end{equation}
    Each noise contribution is plotted in Fig.(\ref{fig:noise}), where we can see the total noise is minimised around $f\sim 10^{-17}-10^{-18}\text{eV}$.    

\section{Sensitivity}
    In this section, we calculate the sensitivity of SOLCN detection to the ALP-photon coupling strength $g_{a\gamma}$. 
    To begin with, we note that for every practical experiment, the real data should be a combination of signal and noise
    \begin{equation}
        \vec{s}(t)=\vec{d}(t)+\vec{n}(t),
    \end{equation}
    where $\vec{s}^{T}(t)=(s_{1}(t),s_{2}(t),\cdots,s_{N_{s}}(t))$ are the data collected by different pairs, and $N_{s}$ is the number of satellite pairs in SOLCN, which in our case is $N_{s}=2$. 
    The correlation between the data from the two pairs is defined as \cite{scpma,sgwb,D59}
    \begin{equation}
        \begin{split}
            C&=\int_{-\frac{\tau}{2}}^{\frac{\tau}{2}}dt\int_{-\frac{\tau}{2}}^{\frac{\tau}{2}}dt's_{1}(t)Q(t-t')s_{2}(t')\\
            &=\int df\int df'\delta_{\tau}(f-f')\tilde{s}_{1}^{*}(f)\tilde{s}_{2}(f')\tilde{Q}(f'),
        \end{split}
    \end{equation}
    where $Q(t-t')$ is a filter function to be determined, $\tau=4\,\text{year}$ is the total observation time and $\delta_{\tau}(f)=\frac{\sin\pi f\tau}{\pi f}$. 
    The signal-to-noise ratio (SNR) is defined as
    \begin{equation}
        \varrho=\frac{\mu}{\sigma},
    \end{equation}
    with
    \begin{equation}
        \begin{split}
            &\mu=\langle C\rangle\\
            &\sigma^{2}=\langle C^{2}\rangle-\langle C\rangle^{2}.
        \end{split}
    \end{equation}
    Using Eq.\eqref{eq:df} and assuming the noise is uncorrelated among the different pairs, we have
    \begin{equation}\label{eq:npsd}
        \begin{split}
            &\langle \tilde{n}_{1}^{*}(f)\tilde{n}_{2}(f)\rangle=0\\
            & \langle \tilde{n}_{i}^{*}(f)\tilde{n}_{i}(f')\rangle=\frac{1}{2}\delta(f-f')S_{n,i}(f),\quad i=1,2
        \end{split}
    \end{equation}
    where we assume in this work that the noise in the two pairs is identical $S_{n,1}(f)=S_{n,2}(f)=S_{n}(f)$. Here $S_{n}(f)$ is the PSD of the noise inside each pair. 
    Substituting Eqs.\eqref{eq:df} and \eqref{eq:npsd} into the expression of the SNR $\varrho$ we obtain
    \begin{equation}
        \begin{split}
            \mu&=\frac{2\tau g_{a\gamma}^{2}\phi_{0}^{2}m_{a}^{2}}{\omega_{0}^{2}}\sin^{2}\left(\frac{m_{a}L}{2}\right)\text{Re}\left(\tilde{Q}\left(\frac{m_{a}}{2\pi}\right)\right)\\
            \sigma^{2}&=\tau\int_{0}^{\infty}df|\tilde{Q}(f)|^{2}S_{n}^{2}(f)
        \end{split}
    \end{equation}
    where we used $\delta_{\tau}(0)=\tau$. 
    Maximizing the SNR we obtain the optimal filter function as
    \begin{equation}
        \tilde{Q}(f)=Q_{0}\frac{\delta(f-\frac{m_{a}}{2\pi})+\delta(f+\frac{m_{a}}{2\pi})}{S_{n}^{2}(\frac{m_{a}}{2\pi})},
    \end{equation}
    where $Q_{0}$ is a normalization constant. 
    Then we obtain the expression for the SNR
    \begin{equation}
        \varrho=
        \frac{g_{a\gamma}^{2}\phi_{0}^{2}m_{a}^{2}}{2\omega_{0}^{2}S_{n}(\frac{m_{a}}{2\pi})}\sin^{2}\left(\frac{m_{a}L}{2}\right)\times
        \begin{cases}
        \quad \tau, \qquad\tau<T_{a}\\
        \sqrt{\tau T_{a}}, \,\quad\tau>T_{a},
        \end{cases}
    \end{equation}
    where $T_{a}=\frac{2\pi}{m_{a}v_{vir}^{2}}$ is the coherence time of the axion halo and the viral velocity is $v_{vir}=10^{-3}$ \cite{rmp93}. 
    Putting numbers into the formula we see that axion with mass $m_{a}<3\times10^{-16}$eV gives $T_{a}>\tau$ while $m_{a}>3\times10^{-16}$eV gives $T_{a}<\tau$.
    
    Setting the threshold for the SNR $\mathcal{\varrho}_{th}=1$ in turn gives the upper bound for the ALP-photon coupling strength
    \begin{equation}\label{eq:gup}
        g_{a\gamma}=
        \frac{\omega_{0}}{\sin(\frac{m_{a}L}{2})}\sqrt{\frac{\varrho_{th}S_{n}(\frac{m_{a}}{2\pi})}{\rho_{a}}}\times
        \begin{cases}
            &\tau^{-1/2},\,\,\,\,\,\qquad\tau<T_{a},\\
            &(\tau T_{a})^{-1/4},\quad\tau>T_{a}.
        \end{cases}
    \end{equation}

\begin{figure}
\begin{center}
\includegraphics[totalheight=55mm]{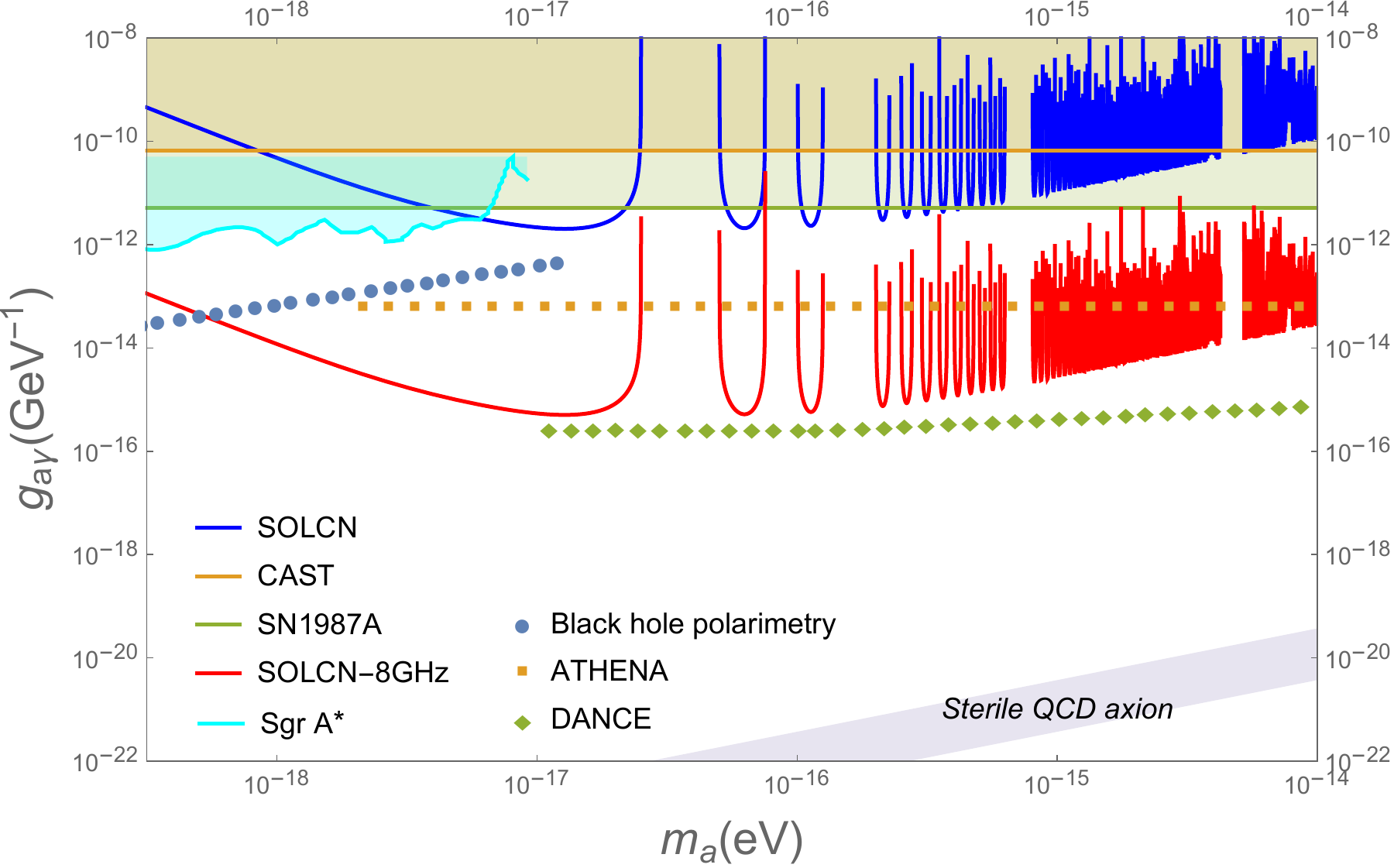}
\caption{Expected sensitivity curves for a space-based optical lattice clock network. 
The solid blue line shows the sensitivity of the SOLCN. 
The solid red line shows the expected sensitivity if we choose an $8$ GHz link frequency, in line with the deep space missions \cite{uop}. 
The shaded area is the region excluded by the present experiments \cite{CAST,sn1987a,sgra}. 
The dotted lines are sensitivity projections from other proposals \cite{bhp,athena,dance}. The sterile QCD axion relation is given in Fig.~6 (right) of Ref.~\cite{Xue2022}.}
\label{fig:sc}
\end{center}
\end{figure}

    Comparing Eq.\eqref{eq:gup} with the constraint for gravitational waves \cite{scpma}, we see a main difference between them is that the sensitivity for the ALPs is crucially reliant on the laser link frequency, while for gravitational waves, the sensitivity is independent of laser link frequency. 
    This property might play a key role in optimizing the ALPs detection sensitivity of SOLCN in the future. 
    
    In Fig.\ref{fig:sc}, we show the expected sensitivity to the ALP-photon coupling strength $g_{a\gamma}$ with the SOLCN, along with relevant constraints from various existing and proposed experiments. 
    For a $430$ THz laser link (corresponding to a wavelength of 698 nm), as in \cite{scpma,D94-12403}, we obtain the solid blue curve indicating a sensitivity of $g_{a\gamma}\geq 10^{-12}\text{GeV}^{-1}$ for ALPs with mass $m_{a}\sim 10^{-17}-10^{-16}\text{eV}$. 
    The result is nearly $1$ order of magnitude below the constraint imposed by the SN1987A observation result \cite{CAST,sn1987a}. 
    
    As mentioned before, the upper bound of $g_{a\gamma}$ relies crucially on the frequency of the laser link. 
    This being the case, a $430$ THz laser link might not be optimal for ALP detection. 
    If instead we use a lower frequency, like the 8 GHz links used in deep space missions, we may significantly increase the sensitivity of the SOLCN \cite{uop,APJ991}.
    
    In the red solid line, we plot the projected sensitivity of the SOLCN with laser link operating at $8\text{GHz}$ \cite{uop}, assuming the noise is maintained at the same order. 
    We show that the SOLCN in this case has the potential to reach $g_{a\gamma}\geq 10^{-16}\text{GeV}^{-1}$ for ALPs with mass $m_{a}\sim 10^{-17}-10^{-16}\text{eV}$. 
    This gives a three orders of magnitude improvement. 

\section{Conclusion and Discussion}
    In this work, we propose ALP detection via synchronized optical lattice clocks.
    By adding a quarter-wave plate along the light path to obtain circularly polarized laser links, ALPs-induced birefringence can be detected via frequency shifts between SOLCN. 
    The motivation for doing so lies in the fact that the ALP field within the targeted mass range is expected to be coherent on the scale of the SOLCN, leading to correlated signals in separated detectors. 
    Thus, the sensitivity can be enhanced by searching for this cross-correlation. 
    
    We have shown in this work the resulting sensitivity curves with two sets of parameters: conservative parameters taken from the stochastic gravitational wave detection proposal in Ref.~\cite{scpma} and optimistic parameters from a optimized scenario with $8$-GHz link frequency. 
    We have demonstrated that the former projection gives nearly one orders of magnitude enhancement in the sensitivity to $g_{a\gamma}$ compared to the SN1987A observation result, while the latter has the potential to reach $g_{a\gamma}\geq 10^{-16}\text{GeV}^{-1}$ for ALPs with masses $m_{a}\sim 10^{-17}-10^{-16}\text{eV}$.

\begin{acknowledgments}
NH is supported by the Beijing Natural Science Foundation (Grant No. IS23025) and the National Natural Science Foundation of China (Grant No. 10056000202402).
\end{acknowledgments}

\bibliography{apssamp}

\end{document}